# Using ZnO-CdS composite nanofibers in the photolytic activity under sunlight irradiation

Mustafa J. Mezher[1], Muhsin A. Kudhier[1] and Osama A. Dakhil[2]

**Corresponding author:** mustafa.j.mezher@uomustansiriyah.edu.iq

**1:** Physics Department, College of Education, Mustansiriyah University, Baghdad, Iraq
**2:** Physics Department, College of Science, Mustansiriyah University, Baghdad, Iraq

## Abstract

In recent years element doping has been extensively employed to modulate the physicochemical properties of semiconductors. In this work, new Pure ZnO and ZnO-xCdS composite nanofibers synthesized through the electrospinning technique, where x= (0.25, 0.5, and 1). Investigated were the effects of doping a ZnO nanofiber composite with varying concentrations of CdS nanoparticles. The crystal structure, morphology, optical properties, and photodegradation activity of Methylene blue (MB) dye were examined in the created nanostructures. XRD pattern reveals that the both ZnO nanofibers and CdS nanoparticles have a mixed phase with a hexagonal wurtzite structure In the ZnO-xCdS composite nanofibers, The effects of composite CdS nanoparticles with ZnO nanofibers were examined using FESEM, the smooth surface of pure ZnO nanofibers, which have a diameter of (33–71) nm and a length of several microns, becomes granular when CdS nanoparticles are added. When ZnO-1CdS nanofibers were compared to pure ZnO nanofibers, the energy band gap, as determined by UV-visible spectroscopy, decreased from 3.79 eV to 2.77 eV. To determine the photocatalytic activities of pure ZnO and ZnO-xCdS composite nanofibers, methylene blue (MB) aqueous solution degradation was used. After 180 minutes, the tests showed that ZnO-1CdS nanofibers had an effective photocatalytic activity of 94.5 percent.

# 1. Introduction

One of the four main worldwide issues is believed to be environmental pollution. Growing industrial development and its usage of organic pollutants exacerbate this issue and can generate a lot of organic pollutants. One of them, methylene blue (MB), is a cationic dye. It is used in numerous industrial applications, including the dyeing of cotton, paper, plastics, and silk [1-4]. We are currently dealing with a serious problem with environmental degradation methods for the elimination of these organic contaminants. The best method for destroying organic pollutants in wastewater is photocatalysis (advanced chemical process), which is both economical and environmentally beneficial [5, 6].

Recently, metal oxide semiconductor nanostructure materials have received extensive attention owing to their distinctive properties, which lead to enhanced physical and chemical properties [7-10]. Semiconductor photocatalysts can be excited by sunlight, which can decompose pollutants in air or water into simple and harmless inorganic compounds, with high efficiency [11-13]. It is well known the use of nano semiconductors like $TiO_2$ and ZnO for this purpose, Because of their exceptional qualities, which include a direct band gap, high electron mobility, the capacity to customize the structures, exciton binding energy, high stability, etc., they have a wide range of possible applications in optics and related fields. Likewise, it is known that the wide band gap of these metal oxides limits their use in the visible range. Additionally, rapid recombination of hole-electrons pairs is another limitation of ZnO, both conditions reduce optical performance of ZnO, which has an impact on the photocatalytic activity. Therefore, the development of new generation nano photocatalysts is a challenge for improving their photocatalytic activity in visible light [5, 14-18].

CdS is a significant semiconductor material with advantageous photoelectric properties. On the one hand, it possesses a narrow band gap, which may contribute to its higher visible photocatalytic activity. However, because CdS itself has metal sulfide properties, its practical use is constrained by photo corrosion and easy oxidation. However, using CdS to support the ZnO nanofiber and boost the composite's photocatalytic activity is still a very efficient technique.[11, 18]. As a result, In addition to being activated by visible light, the ZnO-xCdS composites nanofiber also have a limited capability for photoelectron-hole pair recombination [6].

The relevance of one-dimensional (I-D) nanostructures has increased recently as a result of their novel physical and chemical characteristics and the enhanced photocatalytic activities brought about by their unique geometrical structures. Nanofibers (NFs) have drawn particular attention among ID nanostructures because of their large surface area, high aspect ratio, favorable mechanical characteristics, and outstanding physical characteristics. Additionally, NFS have simple electron transmission as compared to other nanostructures, which can speed up the photodegradation process. Additionally, NFS have simple electron transmission as compared to other nanostructures, which can speed up the photodegradation process [1, 6].

The most popular method for creating nanofibers (NFs) is called electrospinning (ES). Compared to alternative methods for doing so, electrospinning offers the advantages of controllability, simplicity, cheap cost, and manageability in terms of producing large quantities. Additionally, solution and processing conditions affect nanofiber diameter and length [1, 6, 19].

In the present work, samples of easily synthesized ZnO-xCdS composite nanofibers were investigated using XRD, FESEM, and UV-vis spectroscopy. When exposed to sunshine, the photocatalyst exhibited high photoactivity for MB degradation.

## 2. Experimental

### 2.1. Materials

Zinc acetate dihydrate, (Scharlau, Spain Mw=219.49), poly (vinylpyrroridone) (PVP, Mw = 1 300 000), ethanol (98%), ethylene glycol (EG) (Thomas Baker), thiourea (Scharlau, Spain Mw=76.11), and cadmium chloride (Central Drug House CDH, Mw=201.33).

### 2.2. Preparation of ZnO nanofibers

In a typical procedure, a solution was prepared by two steps, first- dissolving 4 g of zinc acetate in 10 ml of distilled water by mixing for 1hour. Second- 10 mL of PVP solution dissolved in ethanol and also mixing for 1hour, now the solution of PVP added to 1 ml of zinc acetate solution. the resulting mixture was continuously stirred for 45 minutes.

The produced solution was immediately added into a 5 mL syringe with 21-gauge needle. The emitting electrode of a power supply of direct current, was attached to the needle. The collection plate, an aluminum plate with clean glass substrates bonded to it, distance between the steel needle tip and the aluminum plate was 20 cm, when a 15 kV voltage was applied between the collective plate and the needle tip, a liquid jet was expelled from the nozzle. A syringe pump was used to regulate the 0.8 ml/h flow rate at which the fluid was given. As the jet moves closer to the collecting plate, The solvent evaporates, leaving the plate with only very thin fibers as seen in (Figure 1). The generated fibers were placed in a humidity bath for around 8 hours to allow for complete hydrolysis of TIP. The remaining PVP material was subsequently eliminated by calcining the fibers for three hours at a temperature of 500°C, at a rate of 5°C/min.

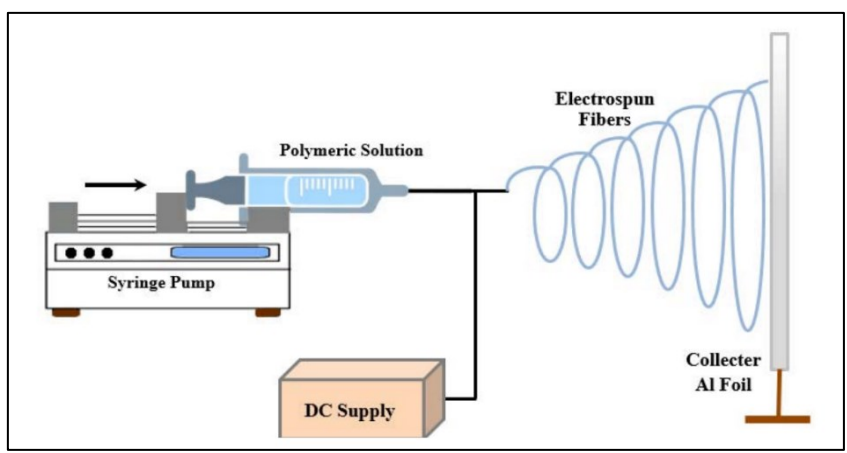

Figure 1 Schematics for the procedure of colloid-electrospinning

2.3. Preparation of ZnO-CdS composite nanofibers

Cadmium chloride CdCl2, which had a concentration of (0.5, 1, 2) g was add in 12.5 ml of ethylene glycol while being constantly stirred at 70–80 ºC. Approximately 0.67, 1.35, and 2.7 g of thiourea that had previously been dissolved in 12.5 ml of ethylene glycol were added drop wise to the solution containing Cd ions. The obtained samples of pure ZnO nanofibers were immersed in Cd solution, the temperature was raised to 100 ºC and maintained there for 2.5 hours. After that, ethanol and water are used to wash the samples.

Finally, Three ZnO-CdS samples were heated for two hours at 300 ºC, following the creation of the samples with various molar concentrations of CdS, they were labeled with ZnO-xCdS, where (x = 0.25, 0.5, and 1).

### 2.4. Characterization Techniques

When creating ZnO and ZnO-xCdS nanofibers, an X-ray diffractometer using CuK1 was utilized to describe the crystalline phase of the nanofibers (U.S. Monoger). Pure ZnO and ZnO-xCdS composite nanofibers' absorption spectra were obtained using a Shimadzu UV-1601 UV-vis spectrophotometer. Additionally, when the samples had been coated in gold, they were examined using a field emission scanning electron microscope (FESEM; ZIESS).

### 2.5. Photocatalytic Activity

The photodegradation of an MB dye aqueous solution in a photochemical container were investigated using pure ZnO and ZnO-xCdS composite nanofibers. An aqueous solution of 0.035 g/L MB dye made with distilled water was used as a model pollutant. was used to study the photocatalytic activity of ZnO-xCdS. 50 mL of MB solution was put to a quartz beaker along with 20 mg of ZnO-xCdS composite nanofibers. The produced samples were mixed with the MB dye solution for 50 min Under dark to create an equilibrium between the adsorption and desorption of the dye molecules on the photocatalyst surface. The suspension was then placed in the sun light after that. To maintain its stability during the experiment, the suspension was gently stirred. A time-dependent fluctuation in the supernatant's absorbance between 550 and 750 nm was discovered using UV-vis analysis after 4 mL of the suspension was taken out of the suspension at 20-minute intervals. The degradation of the MB dye was examined by measuring the difference in the absorption peak at 664 nm with a UV-vis spectrophotometer standard.

## 3. Results and discussion

### 3.1. X-ray diffraction (XRD) analysis

The X-ray diffraction experiments were carried out to determine the crystalline structure of ZnO and ZnO-xCdS composite nanofibers. As shown in Figure 2 The diffracted Bragg peaks were found to match close to the standard ICSD of hexagonal wurtzite structured zinc oxide (Card No: 01-079-0208). In contrast, the other peaks refer to a hexagonal crystal structure of CdS that corresponds to (Card No: 01-080-0006). With the increase in CdS molar concentration from 0.25 to 1, the intensity of hexagonal CdS peaks also increased. The creation of the ZnO-CdS nanocomposites system was confirmed by the appearance of transparent and distinct ZnO and CdS Bragg peaks.

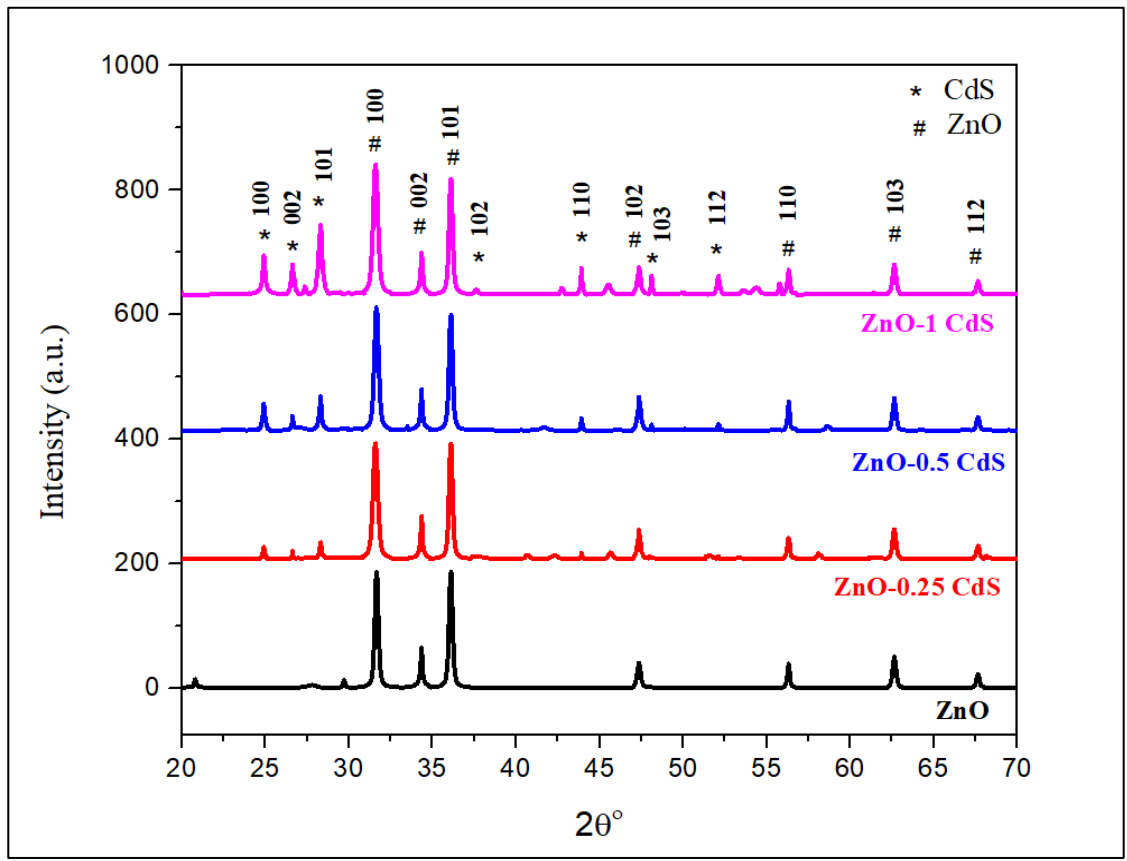

Figure 2: XRD Patterns of pure ZnO and ZnO-xCdS composite nanofibers

3.2. FESEM Study

Pure ZnO and ZnO-xCdS composite nanofibers are displayed in Fig 3 using FESEM images.

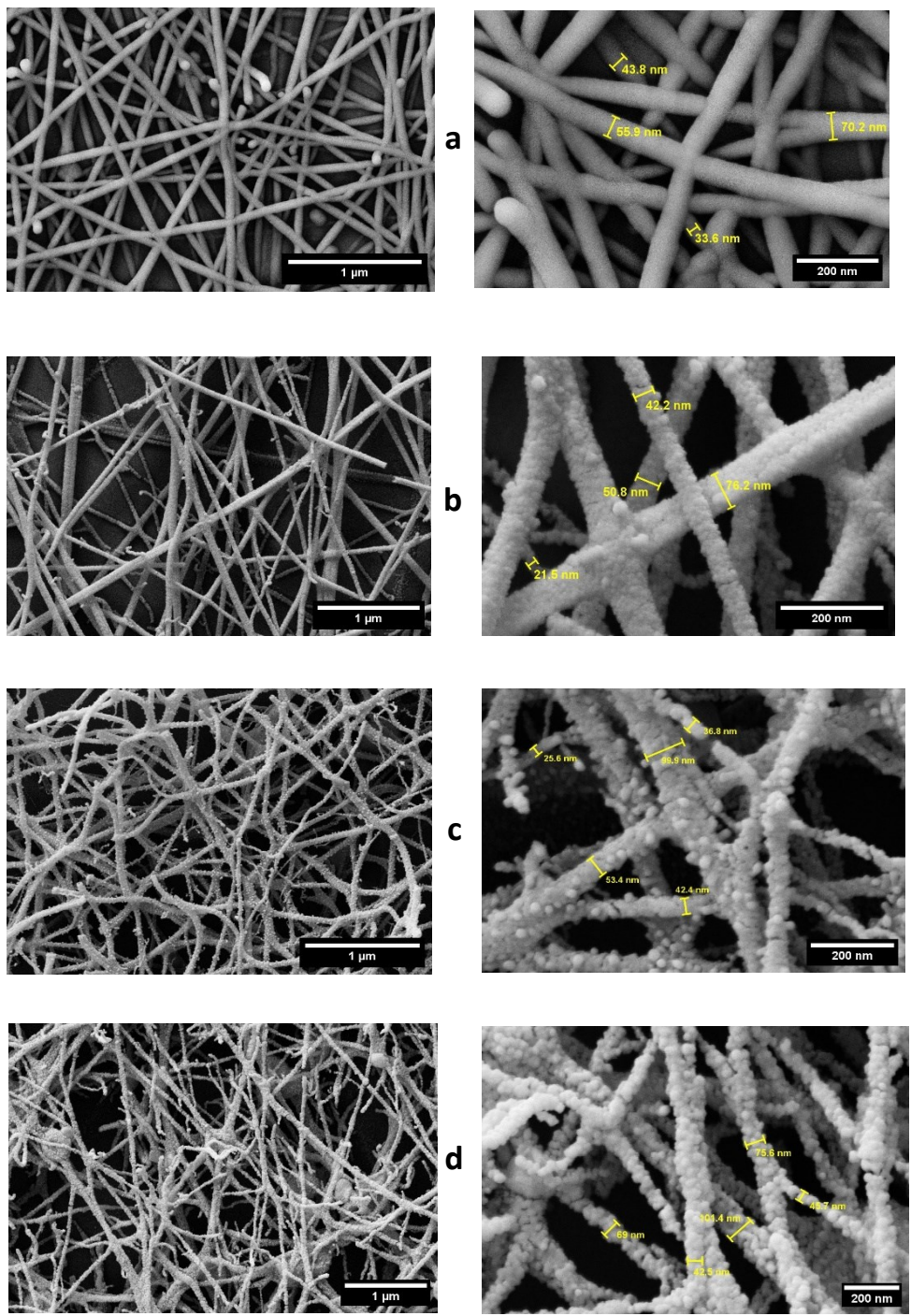

Figure 3: FESEM images of (a) ZnO, (b) ZnO- 0.25CdS, (c) ZnO-0.5CdS, and (d) ZnO-1CdS. Composite nanofibers

Figure 3a shows homogeneous nanofibers of pure ZnO with a length of several microns and a diameter of (33–71) nm. After composite with CdS, the products still have the nanofiber structure, and many CdS nanoparticles are doped to these ZnO nanofibers. Using FESEM images, the stoichiometry of CdS to ZnO nanocomposites and homogeneous surface morphology are studied. These images demonstrate how the addition of CdS nanoparticles causes the smooth surface of pure ZnO nanofibers to convert into a granular surface. This situation is depicted in great detail in Figures 3-(b, c, and d), which also demonstrate how CdS nanoparticles cling to the surface of the nanofibers, increasing the total surface area, which is necessary for the exploitation of photocatalytic activity [20].

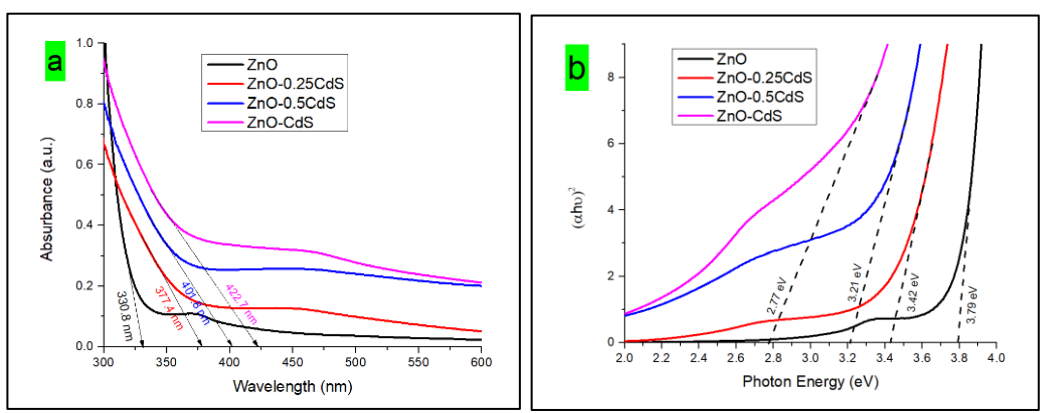

Figure 4: (a) UV–vis absorption spectra of the ZnO-xCdS Composite nanofibers (b) corresponding Tauc plots of $(\alpha h \nu)^2$ versus photon energy of ZnO-xCdS Composite nanofibers

3.3. UV-vis absorbance

The UV-vis absorbance spectra of the ZnO-xCdS Composite nanofibers are shown in Figure 4a. This figure demonstrates how the addition of CdS to ZnO nanofiber increases its solar absorbance, and how the ZnO-xCdS samples' absorption intensities get better as the concentration (x) goes up. This is due to the fact that CdS has better photocatalytic activity than ZnO and a lower energy band gap than ZnO, which increases the samples' absorption intensity. Light with a wavelength below a threshold is absorbed by semiconductor materials (fundamental absorption edge $\lambda_g$)

[21]. Pure ZnO has an absorbing edge below 330.8 nm, whereas ZnO-0.25CdS shifts into the visible range at 377.4 nm, and ZnO-1CdS increases to 422.7 nm. Figure 4b shows tauc graphs derived from absorption spectra. the energy band gap decrease from (3.79, 3.42, 3.21, and 2.77) for (pure ZnO, ZnO -0.25CdS, ZnO -0.5CdS, and ZnO -1CdS) respectively, this decrease in the band gap values indicates that the viewable region is covered by the absorption capacity. These findings showed that the ZnO-xCdS composite nanofibers have the advantage of efficient solar spectrum utilization, resulting in increased photocatalytic efficiency [22, 23].

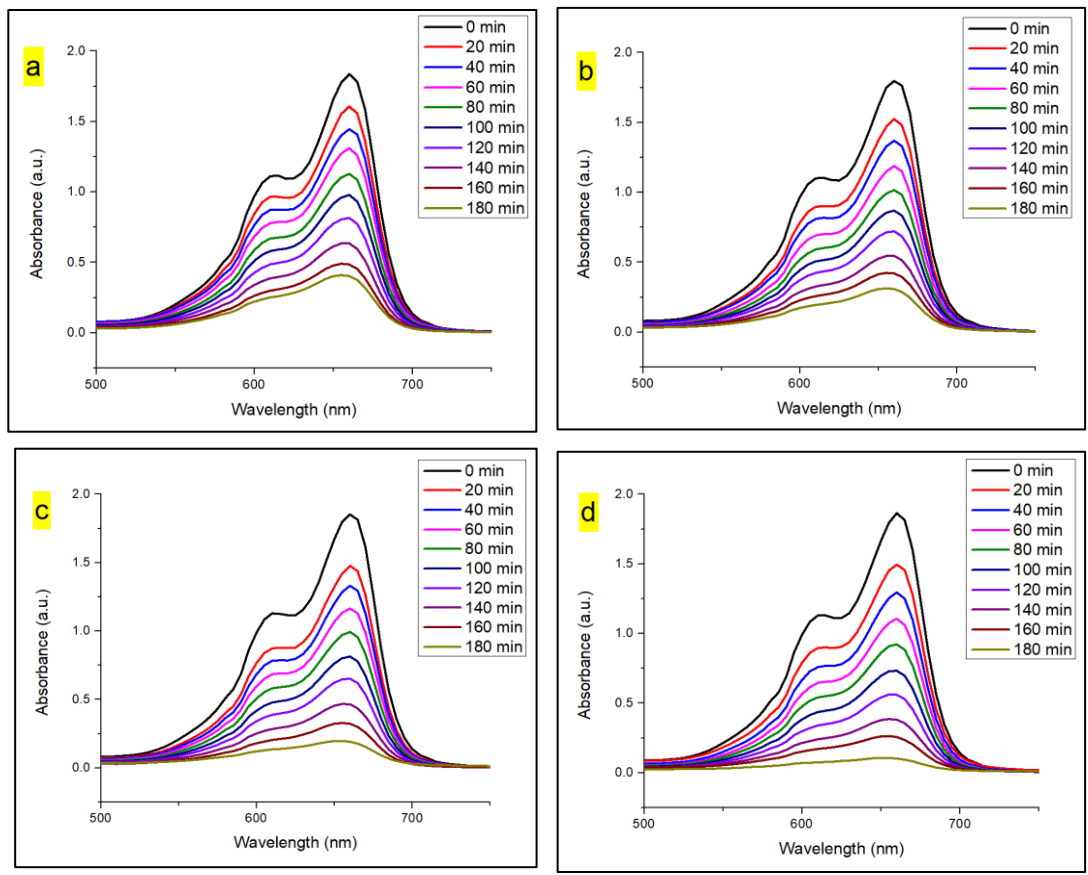

Figure 5: UV-vis absorption spectra of MB dye degradation by (a) pure ZnO, (b) ZnO-0.25CdS, (c) ZnO-0.5CdS, and (d) ZnO-1CdS composite nanofibers.

### 3.4. Photocatalytic Activity

The UV-vis photocatalytic absorption spectra of MB recorded by (ZnO and ZnO-xCdS composite nanofibers) at gradually varying irradiation duration are shown in Figures 5(a), (b), (c), and (d) respectively. The findings clearly show that adding inorganic CdS semiconductors to the ZnO structure can considerably improve the photocatalytic performance of ZnO. When CdS nanoparticles and ZnO nanofibers contact, the electrons at the ZnO-CdS interface would rearrange and migrate from CdS to ZnO, leaving holes in the CdS. The internal electric field at the interface is created as a result of the creation of a depletion layer on the CdS surface and an accumulation layer on the ZnO surface. The electrons were halted from traveling further by the internal electric field that extended from the CdS surface to the ZnO surface. Finally, the electron mobility from CdS to ZnO approaches equilibrium and the two Fermi levels are at the same level [6, 15].

Photocatalytic performance measures how well a catalyst can degrade organic pollutants. The photo-degradation efficiencies of the ZnO and ZnO-CdS composite nanofiber were calculated using Eq. (1) [5], are shown in Figure 6. It can clearly be seen that the ZnO-1CdS nanocomposite sample shows maximum photodegradation efficiency (PE) of almost 95% in 180 min which is much higher in comparison to other samples at same time.

$$\eta\ (\%) = \frac{C_o - C}{C_o} \times 100\% \ \ldots\ldots\ldots\ldots\ldots (1)$$

where $\eta$ is the photodegradation efficiency, $C_o$ is the concentration of MB dye present in the suspension before illumination, and C is the concentration of MB dye present in the suspension at time t. By fitting the photodegradation data from Eq. (2) to a pseudo-first-order kinetic model, the photocatalytic activity was thoroughly evaluated

$$\ln(\frac{C_o}{C}) = kt \ldots\ldots\ldots\ldots(2)$$

Where k is the rate constant [24]. The slope of the line in Figure 7 can be used to determine the rate constant. According to figure 8, the value of k is 0.0083 min$^{-1}$ for pure ZnO and 0.0094, 0.01156, and 0.01416 min$^{-1}$ for nanofibers made of ZnO-0.25CdS, ZnO-0.5CdS, and ZnO-1CdS respectively.

Recent studies show that the ZnO-CdS composite nanofiber outperformed pure ZnO in terms of photocatalytic dye degradation efficiency. In Table 1, a comparison of the photocatalytic effectiveness of ZnO-CdS nanofibers with results from the present [23].

Table 1: Comparison of the ZnO-CdS nanofibers photocatalytic dye degradation efficiency

| Sample | Dye | Light source | Irradiation time (min) | Efficiency (%) | Reference |
|---|---|---|---|---|---|
| CdS/ZnO | MB | Solar light | 240 | 91 | [25] |
| ZnO/CdS | MB | Visible | 120 | 88.5 | [26] |
| CdS/ZnO | MB | Visible | 60 | 85 | [27] |
| CdS/ZnO | MB | Visible | 100 | 90 | [1] |
| ZnO-CdS | MB | Solar light | 180 | 94.5 | Present work |

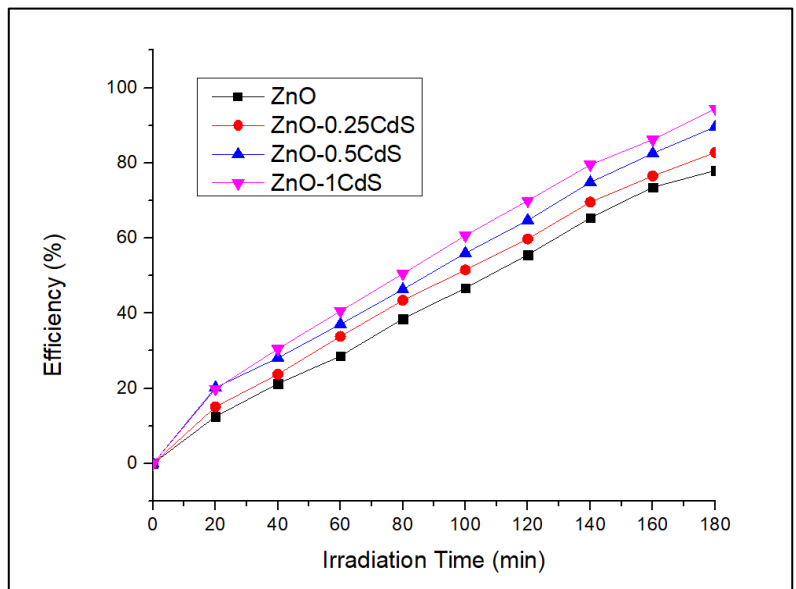

Figure 6: Efficiency of MB dye degradation utilizing of pure ZnO and ZnO-xCdS composite nanofibers.

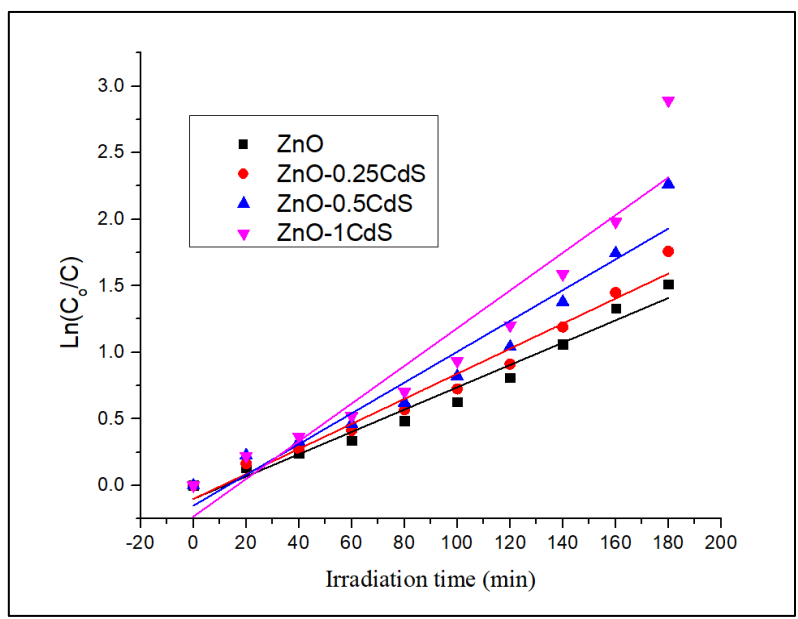

Figure 7: The degradation of MB dye in the first order kinetics as a function of irradiation time of pure ZnO and ZnO-xCdS composite nanofibers.

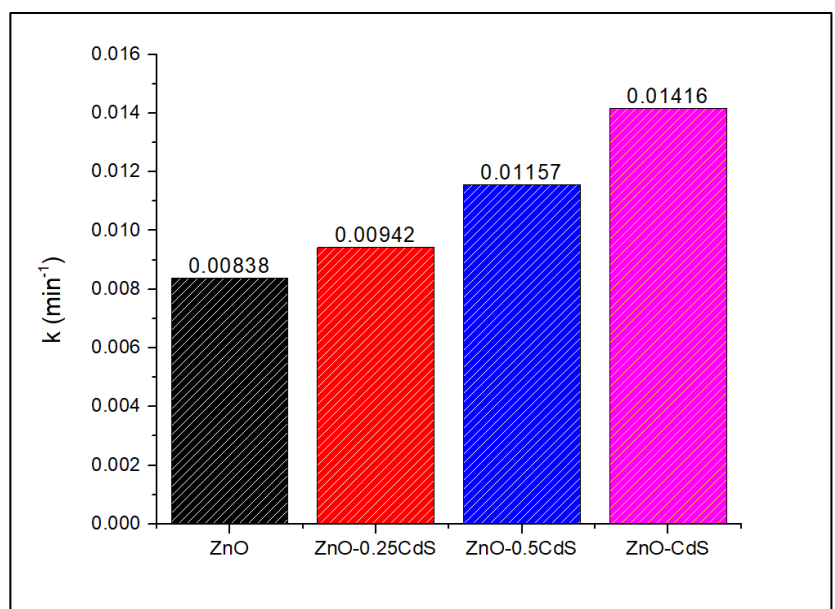

Figure 8: rate constant (k) of pure ZnO and ZnO-xCdS composite nanofibers.

The photocatalytic effectiveness of ZnO-CdS nanocomposites has significantly improved, as shown by photocatalytic degradation results. This enhancement can be due to the following reasons:

(1) As a result of the interaction between the CdS NPs and ZnO NFs, the band gap of ZnO was reduced, as evidenced by the faster transport of charge carriers as a result of the reduced recombination rate of the photogenerated electron-hole pairs, which increased the number of reactive species for the MB degradation.

(2) The composite NFs have a large surface area, which increases the amount of active sites and, in turn, improves how photogenerated charge carriers interact with the molecules that are adsorbed, generating hydroxyl and superoxide radicals that can break down organic molecules [28, 29].

**Conclusion**

Using a simple electrospinning process, we were able to create ZnO-CdS composite nanofibers with a range of CdS loadings from a low-degradability polymer. After calcination, the composite NFs exhibit an excellent nanofibrous structure. The crystalline hexagonal phase of CdS and ZnO makes up the composite NFs. ZnO nanofibers band gap was decreased by the inclusion of CdS nanoparticles. The photocatalytic activity of ZnO nanofibers is significantly enhanced by the addition of CdS nanoparticles. The best photocatalytic activity is shown by ZnO-1CdS composite nanofibers, and after irradiation time 180 minutes, the efficiency approaches 94.5 percent.

**Acknowledgment:** The authors gratefully acknowledge the support they received for this work from the Physics Department, Sciences College, Mustansiriyah University.

**References**

1       Al-Enizi, A.M., El-Halwany, M.M., A. Al-Abdrabalnabi, M., Bakrey, M., Ubaidullah, M., and Yousef, A.: 'Novel Low Temperature Route to Produce CdS/ZnO Composite Nanofibers as Effective Photocatalysts', Catalysts, 2020, 10, (4)
2       Qi, K., Cheng, B., Yu, J., Ho, W.J.J.o.A., and Compounds: 'Review on the improvement of the photocatalytic and antibacterial activities of ZnO', 2017, 727, pp. 792-820

3       Singh, S., and Khare, N.J.C.P.L.: 'CdS/ZnO core/shell nano-heterostructure coupled with reduced graphene oxide towards enhanced photocatalytic activity and photostability', 2015, 634, pp. 140-145

4       Jasim, M.M., Dakhil, O.A.A., and Abdullah, H.I.J.S.S.S.: 'Synthesis of NiO/TNTs pn junction for highly photocatalysis activity under sunlight irradiation', 2020, 107, pp. 106342

5       Saxena, N., Sondhi, H., Sharma, R., Joshi, M., Amirthapandian, S., Rajput, P., Sinha, O.P., and Krishna, R.J.C.P.I.: 'Equimolar ZnO-CdS nanocomposite for enhanced photocatalytic performance', 2022, pp. 100119

6       Kudhier, M.A., ALKareem, R.A.S.A., and Sabry, R.S.J.J.o.t.M.B.o.M.: 'Enhanced photocatalytic activity of TiO2-CdS composite nanofibers under sunlight irradiation', 2021, 30, (1), pp. 213-219

7       Das, A., Nikhil, S., Nair, R.G.J.N.-S., and Nano-Objects: 'Influence of surface morphology on photocatalytic performance of zinc oxide: A review', 2019, 19, pp. 100353

8       Chaudhary, S., Kaur, Y., Jayee, B., Chaudhary, G.R., and Umar, A.J.J.o.C.P.: 'NiO nanodisks: Highly efficient visible-light driven photocatalyst, potential scaffold for seed germination of Vigna Radiata and antibacterial properties', 2018, 190, pp. 563-576

9       Umar, A., Akhtar, M., Al-Hajry, A., Al-Assiri, M., Dar, G., and Islam, M.S.J.C.E.J.: 'Enhanced photocatalytic degradation of harmful dye and phenyl hydrazine chemical sensing using ZnO nanourchins', 2015, 262, pp. 588-596

10      Cheng, X., Liu, H., Yu, X., Chen, Q., Li, J., Wang, P., Umar, A., and Wang, Q.J.S.o.A.M.: 'Preparation of highly ordered TiO2 nanotube array photoelectrode for the photoelectrocatalytic degradation of methyl blue: activity and mechanism study', 2013, 5, (11), pp. 1563-1570

11      Zhou, Q., Li, L., Xin, Z., Yu, Y., Wang, L., Zhang, W.J.J.o.A., and Compounds: 'Visible light response and heterostructure of composite CdS@ ZnS–ZnO to enhance its photocatalytic activity', 2020, 813, pp. 152190

12      Farahi, E., and Memarian, N.J.C.P.L.: 'Nanostructured nickel phosphide as an efficient photocatalyst: effect of phase on physical properties and dye degradation', 2019, 730, pp. 478-484

13      Wang, Z., Li, C., and Domen, K.J.C.S.R.: 'Recent developments in heterogeneous photocatalysts for solar-driven overall water splitting', 2019, 48, (7), pp. 2109-2125

14      Xu, T., Zhang, L., Cheng, H., and Zhu, Y.J.A.C.B.E.: 'Significantly enhanced photocatalytic performance of ZnO via graphene hybridization and the mechanism study', 2011, 101, (3-4), pp. 382-387

15      Wang, S., Zhu, B., Liu, M., Zhang, L., Yu, J., and Zhou, M.J.A.C.B.E.: 'Direct Z-scheme ZnO/CdS hierarchical photocatalyst for enhanced photocatalytic H2-production activity', 2019, 243, pp. 19-26

16      Kegel, J., Povey, I.M., and Pemble, M.E.J.N.e.: 'Zinc oxide for solar water splitting: A brief review of the material's challenges and associated opportunities', 2018, 54, pp. 409-428

17      Zhang, C., Li, N., Chen, D., Xu, Q., Li, H., He, J., Lu, J.J.J.o.A., and Compounds: 'The ultrasonic-induced-piezoelectric enhanced photocatalytic performance of ZnO/CdS nanofibers for degradation of bisphenol A', 2021, 885, pp. 160987

18      Mahala, C., Sharma, M.D., and Basu, M.J.N.J.o.C.: 'ZnO@ CdS heterostructures: an efficient photoanode for photoelectrochemical water splitting', 2019, 43, (18), pp. 7001-7010

19      Su, Y., Lu, B., Xie, Y., Ma, Z., Liu, L., Zhao, H., Zhang, J., Duan, H., Zhang, H., and Li, J.J.N.: 'Temperature effect on electrospinning of nanobelts: the case of hafnium oxide', 2011, 22, (28), pp. 285609

20      Xu, Z., Quintanilla, M., Vetrone, F., Govorov, A.O., Chaker, M., and Ma, D.J.A.F.M.: 'Harvesting lost photons: plasmon and upconversion enhanced broadband photocatalytic activity in core@ shell microspheres based on lanthanide-doped NaYF4, TiO2, and Au', 2015, 25, (20), pp. 2950-2960

21 Usubharatana, P., McMartin, D., Veawab, A., Tontiwachwuthikul, P.J.I., and research, e.c.: 'Photocatalytic process for CO2 emission reduction from industrial flue gas streams', 2006, 45, (8), pp. 2558-2568

22 Mo, Z., Huang, Y., Lu, S., Fu, Y., Shen, X., and He, H.J.O.: 'Growth of ZnO nanowires and their applications for CdS quantum dots sensitized solar cells', 2017, 149, pp. 63-68

23 Gurugubelli, T.R., Ravikumar, R., and Koutavarapu, R.J.C.: 'Enhanced Photocatalytic Activity of ZnO–CdS Composite Nanostructures towards the Degradation of Rhodamine B under Solar Light', 2022, 12, (1), pp. 84

24 Koutavarapu, R., Tamtam, M.R., Lee, S.-G., Rao, M., Lee, D.-Y., and Shim, J.J.J.o.E.C.E.: 'Synthesis of 2D NiFe2O4 nanoplates/2D Bi2WO6 nanoflakes heterostructure: An enhanced Z-scheme charge transfer and separation for visible-light-driven photocatalytic degradation of toxic pollutants', 2021, 9, (5), pp. 105893

25 Velanganni, S., Pravinraj, S., Immanuel, P., and Thiruneelakandan, R.J.P.B.C.M.: 'Nanostructure CdS/ZnO heterojunction configuration for photocatalytic degradation of Methylene blue', 2018, 534, pp. 56-62

26 Nandi, P., Das, D.J.J.o.P., and Solids, C.o.: 'ZnO/CdS/CuS heterostructure: A suitable candidate for applications in visible-light photocatalysis', 2022, 160, pp. 110344

27 Bai, L., Li, S., Ding, Z., Wang, X.J.C., Physicochemical, S.A., and Aspects, E.: 'Wet chemical synthesis of CdS/ZnO nanoparticle/nanorod hetero-structure for enhanced visible light disposal of Cr (VI) and methylene blue', 2020, 607, pp. 125489

28 Yousef, A., Barakat, N.A., Amna, T., Al-Deyab, S.S., Hassan, M.S., Abdel-Hay, A., and Kim, H.Y.J.C.I.: 'Inactivation of pathogenic Klebsiella pneumoniae by CuO/TiO2 nanofibers: A multifunctional nanomaterial via one-step electrospinning', 2012, 38, (6), pp. 4525-4532

29 Yousef, A., Barakat, N.A., Amna, T., Unnithan, A.R., Al-Deyab, S.S., and Kim, H.Y.J.J.o.L.: 'Influence of CdO-doping on the photoluminescence properties of ZnO nanofibers: effective visible light photocatalyst for waste water treatment', 2012, 132, (7), pp. 1668-1677